\documentclass[a4paper]{PoS}

\title{Rapidity profiles from 3+1D Glasma simulations with finite longitudinal thickness}

\ShortTitle{Rapidity profiles from 3+1D Glasma simulations with finite longitudinal thickness}

\author{Andreas Ipp and \speaker{David M\"uller}\\
Institute for Theoretical Physics, TU Wien,\\
Wiedner Hauptstra{\ss}e 8-10, A-1040 Vienna, Austria\\
E-mail: \email{ipp@hep.itp.tuwien.ac.at},
\email{david.mueller@tuwien.ac.at}}

\abstract{
We present our progress on simulating the Glasma in the early stages of heavy ion collisions in a non-boost-invariant setting. Our approach allows us to describe colliding nuclei with finite longitudinal width by extending the McLerran-Venugopalan model to include a parameter for the Lorentz-contracted but finite extent of the nucleus in the beam direction. We determine the rapidity profile of the Glasma energy density, which shows strong deviations from the boost invariant result. Both broad and narrow profiles can be produced by varying the initial conditions. We find reasonable agreement when we compare the results to rapidity profiles of measured pion multiplicities from RHIC.
}

\FullConference{The European Physical Society Conference on High Energy Physics\\
                 5-12 July\\
                 Venice, Italy}

\begin{document}

\begin{figure}
\begin{centering}
\includegraphics
[height=5cm, keepaspectratio]
{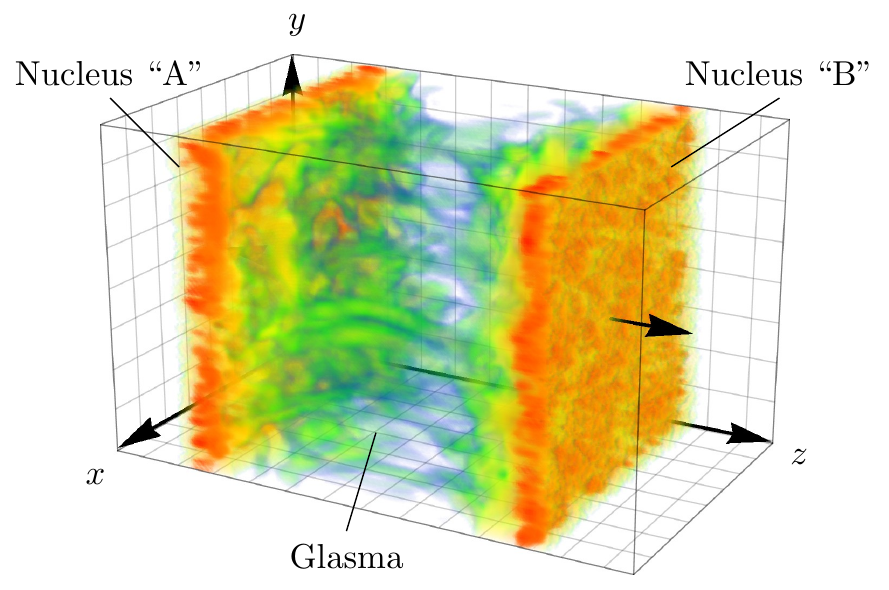}
\par\end{centering}
\caption{
A 3D plot of the energy density directly after the collision, showing both nuclei
``A'' and ``B'' and the Glasma with visible flux tube structure \cite{Ipp:2017lho}.
The box only covers a small part of the full collision in the transverse plane spanned by $x$ and $y$,
for which we use periodic boundary conditions as opposed to fixed boundary conditions
in the longitudinal direction.
\label{fig:fig1}}
\end{figure}

\section{Introduction}

In the earliest stages of heavy ion collisions the color glass condensate (CGC)
effective field theory \cite{Gelis:2012ri, Gelis:2010nm} provides a first principles, weak coupling,
yet non-perturbative description in terms of classical Yang-Mills fields
and color currents.
The main idea is a separation of degrees of freedom in a fast nucleus into 
partons with large longitudinal momentum fraction $x$ (e.g. valence quarks) and
partons with small $x$ (mostly soft gluons).
Large $x$ partons are assumed to be randomly distributed, recoilless classical
color charges, moving at the speed of light, while small $x$ partons are described
as classical gluon fields sourced by the large $x$ partons.
In the ultrarelativistic limit the classical field of the soft gluons becomes a 
shock wave.
The result of a collision of two such shock waves is a field initially composed of
purely longitudinal color flux tubes known as the Glasma \cite{Lappi:2006fp}, a precursor to the
quark gluon plasma.
Due to the assumption of infinitely thin shock waves the Glasma is
boost invariant, i.e. there is no dependence on space-time rapidity
$\eta_{s}$.
After the collision the Glasma evolves and expands classically
according to the Yang-Mills equations, preserving the initial boost
invariance.
This classical evolution only holds for a short amount of time, up to roughly
$\tau\lesssim1\,\mbox{fm}/c$, after which the field quickly becomes decoherent
and the classical approach is considered to be invalid.

The evolution of the boost invariant Glasma and its properties
have been investigated thoroughly both numerically \cite{Krasnitz:1998ns, Lappi:2003bi, Lappi:2006hq, Dumitru:2013koh}
and analytically in the weak field limit \cite{Fujii:2008dd} or using the small $\tau$ expansion
\cite{Chen:2015wia}.
The Glasma can also be studied in settings without boost invariance, for instance
by introducing small, rapidity dependent fluctuations  in the initial conditions 
\cite{Romatschke:2006nk, Fukushima:2011nq}, by considering the 
saturation scale as a function of rapidity \cite{Lappi:2004sf} or more fundamentally at 
next-to-leading order in the gauge coupling, by making use of the JIMWLK
evolution \cite{Iancu:2000hn, Weigert:2005us, Schenke:2016ksl}.
However, in all of these approaches one still uses the Glasma initial conditions
\cite{Kovner:1995ja}, which are derived under the assumption of boost invariance and are thus 
not strictly valid at finite collision energy.

Obviously, realistic nuclei at finite collision energies are not infinitely
thin.
The finite ``classical'' longitudinal extent of a nucleus at a given velocity $v$
is proportional to $R/\gamma$, where $R$ is the radius of the nucleus and
$\gamma^{-1}=\sqrt{1-(v/c)^{2}}$ is the Lorentz factor.
In the context of the JIMWLK evolution, the longitudinal extent of nuclei might
even be larger than the classical extent:
as the evolution to lower longitudinal momenta (smaller $x$) progresses, more and
more small $x$ gluons are added to the gluon field of the nucleus.
Due to uncertainty in the longitudinal position at small $x$, these gluons are
spread out over a length typically larger than $R/\gamma$, which leads to a
picture of rather thick nuclei even at ultrarelativistic energies \cite{Iancu:2011nj}.
Except for a few pioneering studies \cite{Poschl:1998px, Ozonder:2013moa}, the effects of finite longitudinal
extent and its consequences on the evolution of the Glasma have been
largely ignored in the past.

The aim of this work is to extend the CGC/Glasma description to
include the finite longitudinal extent of the colliding nuclei, striving
for a more realistic picture at finite collision energies. In this
proceedings contribution we present our progress in simulating heavy-ion
collisions in the CGC framework at finite collision energies based
on our previous publications \cite{Gelfand:2016yho, Ipp:2017lho}.

\section{Initial conditions}

In the standard picture of a boost invariant collision the color current of a
left moving nucleus (here denoted as ``A'') can be written as 
\begin{equation}
J_{(A)}^{-}(x^{+},x_{T})=\delta(x^{+})\rho_{(A)}(x_{T})\,,
\label{eq:bi_current}
\end{equation}
where  $x^\pm \equiv (x^0 \pm x^3) / \sqrt{2}$ and $\rho_{(A)}(x_{T})$ is the color charge density in the transverse
plane.
Supplied with the covariant gauge condition $\partial_{\mu}A^{\mu}(x)=0$,
the current generates a field given by
\begin{equation}
A_{(A)}^{-}(x^{+},x_{T})=-\Delta_{T}^{-1}J_{(A)}^{-}(x^{+},x_{T})\,,
\label{eq:poisson}
\end{equation}
with the Laplace operator in the transverse plane $\Delta_{T}$.
Likewise, a right moving nucleus (``B'') is described by the color current
$J_{(B)}^{+}(x^{-},x_{T})$ and the field $A_{(B)}^{+}(x^{-},x_{T})$.
As mentioned in the introduction, collisions of two such fields leads to the
boost invariant Glasma picture.
In order to describe the non-boost-invariant scenario where nuclei
have finite longitudinal extent we have to relax the assumption made
in Eq.\,(\ref{eq:bi_current}) and account for extended support of
the color charges along the longitudinal direction. The solution to
the field equations before the collision in Eq.\,(\ref{eq:poisson}) remain
unchanged.
In particular we look at initial conditions where the longitudinal shape
can be separated from the transverse color charge density
\begin{equation}
J_{(A)}^{-}(x^{+},x_{T})=f(x^{+})\rho_{(A)}(x_{T})\,,
\label{eq:current_smear}
\end{equation}
where $f(x^{+})$ is a normalized function that defines the longitudinal profile.
The form of the current is a special case where
non-trivial color structure in the longitudinal extent is neglected \cite{Fukushima:2007ki}.
With extended currents and fields it is no longer possible to just focus on the
forward light cone.
Instead, we solve the full Yang-Mills equations
\begin{equation}
D_{\mu}F^{\mu\nu}(x)=J_{(A)}^{\nu}(x)+J_{(B)}^{\nu}(x)
\end{equation}
in the laboratory frame in 3+1 dimensions.
Working the lab frame forces us to explicitly include the color currents in our
simulation and we have to make sure that the non-Abelian charge conservation
$D_{\mu}J_{(A,B)}^{\mu}(x)=0$ holds.
Details of how we do this numerically are described in \cite{Gelfand:2016yho}.
As a model for large nuclei we use the McLerran-Venugopalan
model \cite{MV1} defined by the charge density correlator
\begin{equation}
\left\langle \rho_{(A,B)}^{a}(x_{T})\rho_{(A,B)}^{b}(y_{T})\right\rangle = 
g^{2}\mu^{2}\delta^{(2)}(x_{T}-y_{T})\delta^{ab}\,,
\label{mv}
\end{equation}
where $g$ is the gauge coupling constant and $\mu$ controls the
average charge fluctuations in the transverse plane.
In our approach (Eq.\,(\ref{eq:current_smear})) the two-dimensional charge density
given by Eq.\,(\ref{mv}) is simply extended in the longitudinal direction.
We relate the width of $f(x^{+})$ (which we take to be a Gaussian) to the Lorentz
contracted diameter of the nucleus $2R/\gamma$, which in turn is related to the
collision energy $\sqrt{s_{NN}}$.
In our most recent publication \cite{Ipp:2017lho} we look at RHIC-like
scenarios of central Au+Au collisions with $\sqrt{s_{NN}}$ at $200\,\mbox{GeV}$ and
$130\,\mbox{GeV}$.

\begin{figure}[t!]
\begin{centering}
\includegraphics{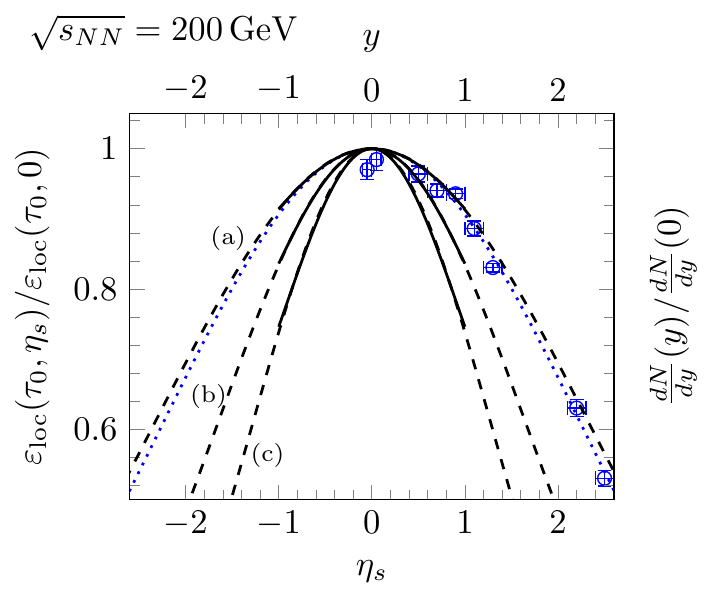}
\includegraphics{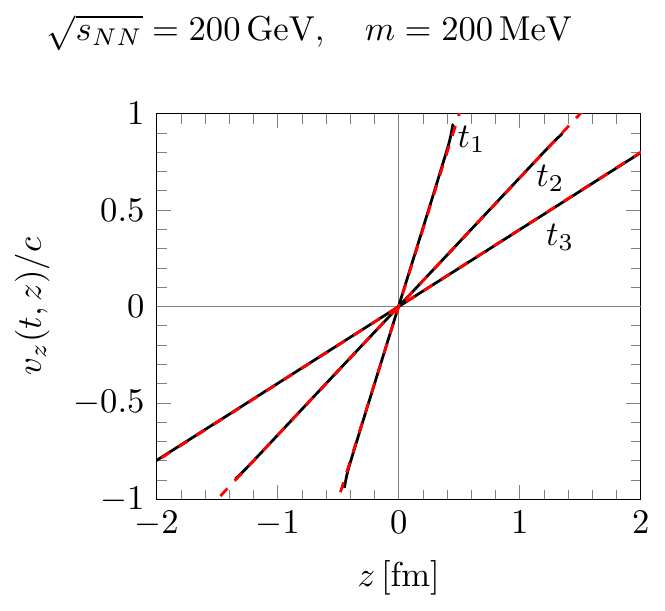}
\par\end{centering}
\caption{
Left:
Comparison of the space-time rapidity profile of the LRF
energy density $\varepsilon_{loc}(\tau_{0},\eta_{s})$ (thick solid lines),
$\pi^{+}$ multiplicity $dN/dy$ at RHIC \cite{Bearden:2004yx} (data points) and Gaussian fits 
(dashed and dotted). The infrared regulator $m$ has a large effect on the width:
(a) $m=0.2\,\mbox{GeV}$, (b) $m=0.4\,\mbox{GeV}$ and (c) $m=0.8\,\mbox{GeV}$.
Right:
Longitudinal velocity $v_{z}(t,z)$ as a function of the longitudinal coordinate
$z$ evaluated at different laboratory frame times $t$ from out simulation 
(black, solid lines) compared to the free-streaming case $v_{z}(t,z) = z/t$
(red, dashed lines). We show three different times:
$t_1 = 0.5\,\mbox{fm}/c$,
$t_2 = 1.5\,\mbox{fm}/c$ and
$t_3 = 2.5\,\mbox{fm}/c$.
\label{fig:fig2}
}
\end{figure}

\section{Results}

Solving the field equations for the collision scenario in 3+1 we obtain pictures
like Fig.\,\ref{fig:fig1}, where we plot the energy density of the Yang-Mills
fields directly after the collision.
We study the Glasma in our simulations by looking at the local rest frame (LRF)
energy density $\varepsilon_{loc}(\tau, \eta_s)$ which we compute by
diagonalizing the energy-momentum tensor $T_{\nu}^{\mu}(x)$.
In Fig.\,\ref{fig:fig2} (left) we show $\varepsilon_{loc}(\tau_0, \eta_s)$ as a function of
space-time rapidity at $\tau_{0}=1\,\mbox{fm}/c$ for $\sqrt{s_{NN}}=200\,\mbox{GeV}$.
We observe that these rapidity profiles are approximately Gaussian in shape.
The width of the profiles depends on the energy $\sqrt{s_{NN}}$ and also
strongly on the infrared regulator that is used in the initial conditions.
As one should expect, we observe that reducing $\sqrt{s_{NN}}$, thus increasing
the longitudinal extent,  the profiles become more narrow.
The effect of the infrared regulator is more surprising, not well understood and
warrants further in-depth studies.
The width itself is rather independent of evaluation time $\tau_{0}$ as
long as $\tau_{0}\gtrsim0.3\,\mbox{fm}/c$ after which one enters the
free-streaming limit.

Free-streaming can be observed by looking at the longitudinal velocity $v_{z}$,
which corresponds to the velocity associated with the Lorentz
boost that transforms from the laboratory frame into the LRF.
In Fig.\,\ref{fig:fig2} (right) we plot $v_{z}(t,z)$ as a function of the
longitudinal coordinate $z$ at different times $t$ and compare it
to the free-streaming case, where $v_{z}=z/t$.
We observe that the two curves match, which implies that the LRF
mostly corresponds to the $(\tau,\eta_{s})$ frame.
The free-streaming limit is also visible in the strong pressure anisotropy at
later times where transverse pressure $p_{T}$ dominates longitudinal
pressure $p_{L}$.

The conclusion we can draw from these observations is that by including
finite longitudinal extent, thus explicitly breaking the boost invariance
of the system, we obtain non-flat rapidity profiles that develop early
on in the evolution, but the Glasma still flows in a free-streaming
manner just like in the boost invariant approximation.
Finite longitudinal extent therefore does not fundamentally change the picture of
the boost invariant Glasma, except for non-flat rapidity profiles.

\acknowledgments
The authors thank A. Kurkela and T. Lappi for helpful discussions.
This work has been supported by the Austrian Science
Fund FWF, Project No.\,P26582-N27 and Doctoral
program No.\,W1252-N27. The computational results
have been achieved using the Vienna Scientific Cluster.

\bibliographystyle{JHEP}
\bibliography{references}

\end{document}